\documentclass[usenatbib]{mn2e}

\pdfoutput=1
\usepackage[english]{babel}
\usepackage[babel]{csquotes}
\usepackage{indentfirst}
\usepackage{setspace}
\usepackage{fancyhdr}
\usepackage{graphicx}
\usepackage{booktabs}
\usepackage{amsmath}
\usepackage{tensor}
\usepackage{placeins}
\usepackage{natbibmnfix}
\usepackage{epsfig}
\usepackage{aas_macros}
\usepackage{color}
\usepackage{url}

\def\gtsima{$\; \buildrel > \over \sim \;$} 
\def\ltsima{$\; \buildrel < \over \sim \;$} 
\def\gsim{\lower.5ex\hbox{\gtsima}} 
\def\lsim{\lower.5ex\hbox{\ltsima}} 
\def\simgt{\lower.5ex\hbox{\gtsima}} 
\def\simlt{\lower.5ex\hbox{\ltsima}} 
 
\def\Lya{Ly$\alpha$~}

\def\lya { {Ly\alpha} }

\def\cross{\psi_{\alpha L}}

\title[Empowering intensity mapping]{Empowering line intensity mapping to study early galaxies}
\author[P. Comaschi, A. Ferrara] {P.~Comaschi$^1$\thanks{Email: paolo.comaschi@sns.it}, A.~Ferrara$^{1,2}$\\
$^1$Scuola Normale Superiore, Piazza dei Cavalieri 7, 1-56126 Pisa, Italy\\
$^2$Kavli IPMU, The University of Tokyo, 5-1-5 Kashiwanoha, Kashiwa 277-8583, Japan}
\date{March 28, 2016}

\def\LaTeX{L\kern-.36em\raise.3ex\hbox{a}\kern-.15em
    T\kern-.1667em\lower.7ex\hbox{E}\kern-.125emX}

\begin{document}
	
	\maketitle

\begin{abstract}
Line intensity mapping is a superb tool to study the collective radiation from early galaxies. However, the method is hampered by the presence of strong foregrounds, mostly produced by low-redshift interloping lines. We present here a general method to overcome this problem which is robust against foreground residual noise and based on the cross-correlation function $\cross(r)$ between diffuse line emission and \Lya emitters (LAE). We compute the diffuse line (\Lya is used as an example)  emission from galaxies in a $(800{\rm Mpc})^3$ box at $z = 5.7$ and $6.6$. We divide the box in slices and populate them with $14000(5500)$ LAEs at $z = 5.7(6.6)$, considering duty cycles from $10^{-3}$ to $1$. Both the LAE number density and slice volume are consistent with the expected outcome of the Subaru HSC survey. We add gaussian random noise with variance $\sigma_{\rm N}$ up to 100 times the variance of the \Lya emission, $\sigma_\alpha$, to simulate foregrounds and compute $\cross(r)$.
We find that the signal-to-noise of the observed $\cross(r)$ does not change significantly if $\sigma_{\rm N} \le 10 \sigma_\alpha$ and show that in these conditions the mean line intensity, $I_\lya$, can be precisely recovered independently of the LAE duty cycle. Even if $\sigma_{\rm N} = 100 \sigma_\alpha$, $I_\alpha$ can be constrained within a factor $2$. The method works equally well for any other line (e.g. HI 21 cm, [CII], HeII) used for the intensity mapping experiment.
\end{abstract}


\section{Introduction}
\label{sec:intro}
The Epoch of Reionization (EoR,  redshift $z\simgt 5.5$) is a key phase of cosmic evolution for a number of reasons: (i) it is the last major cosmic phase transition \citep{2001PhR...349..125B}; (ii) it affected a vast majority of the baryons in the universe; (iii) it has been at the edge of our observational capabilities for many years \citep{2014ApJ...793..115B, 2014ApJ...786..108O, 2015ApJ...804L..30O, 2010ApJ...723..869O, 2008ApJS..176..301O, 2015MNRAS.451..400M}. Most importantly, it is intimately connected with the formation of the first galaxies and to the very beginning of the universe as we know it. 

Unfortunately its observational investigation has proven extremely challenging. EoR galaxies are extremely faint and therefore hard to detect. Moreover the most popular theoretical scenario foresees that reionization is powered by a population of low mass, numerous galaxies \citep{2011MNRAS.414..847S} whose cumulative emission, dominates over more massive outliers. These ``typical'' high redshift galaxies are individually out of reach of even the most powerful observatories of the next decades (such as JWST\footnote{\url{http://www.jwst.nasa.gov}} ,TMT\footnote{\url{http://www.tmt.org}} or ELT\footnote{\url{https://www.eso.org/sci/facilities/eelt/}}).

These difficulties make the development of new observational strategies targeting the cumulative emission of galaxies \citep{2005PhR...409..361K, 2016arXiv160203512C}, rather than individual detection of sources, quite compelling. Intensity mapping (IM) is one implementation of this approach and targets the 3D fluctuations in the large scale cumulative line emission \citep{2008A&A...489..489R, 2010JCAP...11..016V, 2011JCAP...08..010V}. In principle, it can be applied independently of the particular line chosen, and several alternatives have been proposed in recent years (e. g. HI 21cm \citep{2006PhR...433..181F}, CO \citep{2011ApJ...741...70L, 2008A&A...489..489R, 2014MNRAS.443.3506B}, [CII] \citep{2012ApJ...745...49G, 2014arXiv1410.4808S, mappingCII}, H$_2$ \citep{2013ApJ...768..130G}, HeII \citep{2015arXiv150103177V} and \Lya \citep{2014ApJ...786..111P, 2013ApJ...763..132S, 2016MNRAS.455..725C}.
\begin{figure*}
\vspace{+0\baselineskip}
{
\includegraphics[width=0.45\textwidth]{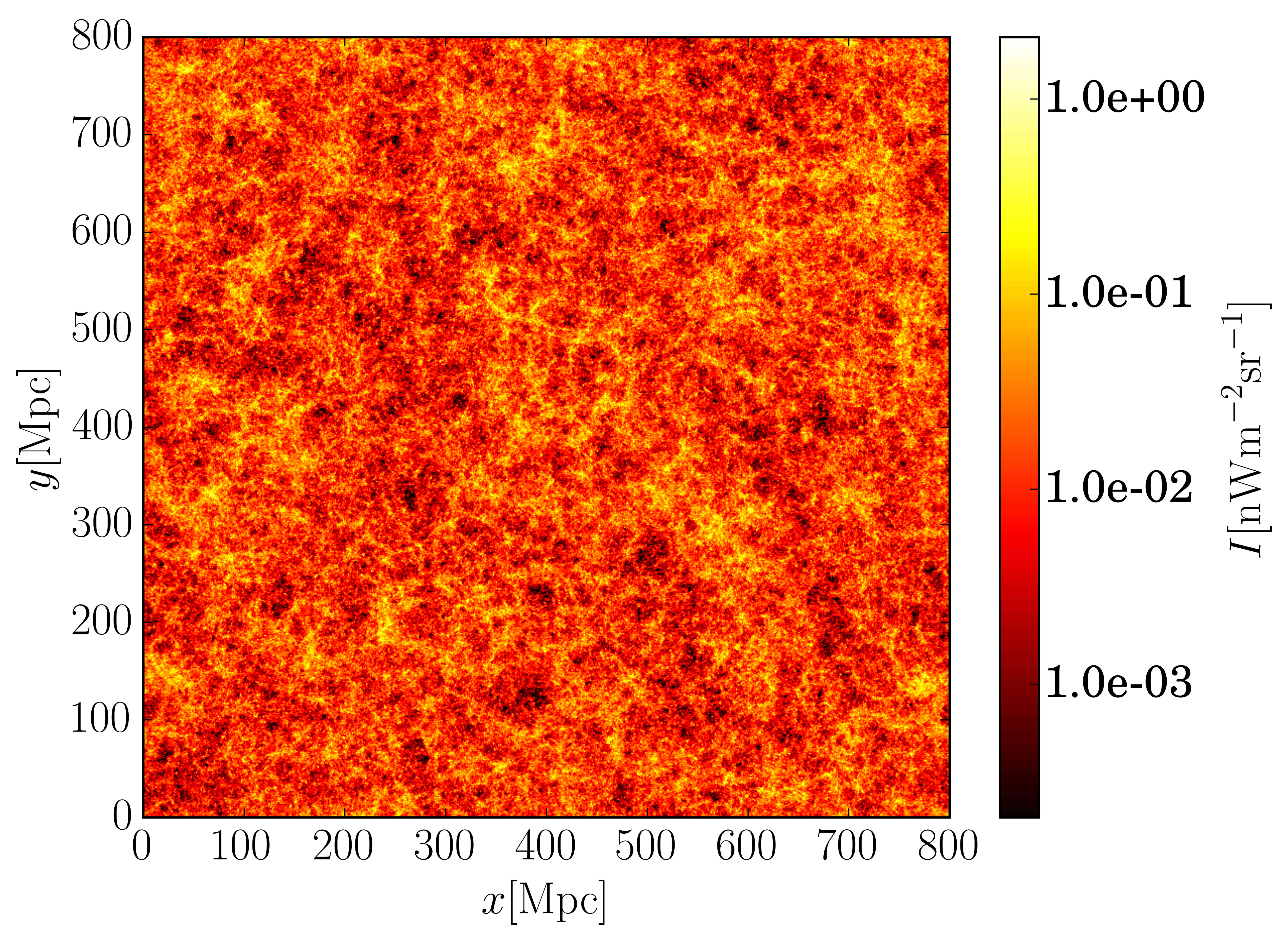}
\includegraphics[width=0.45\textwidth]{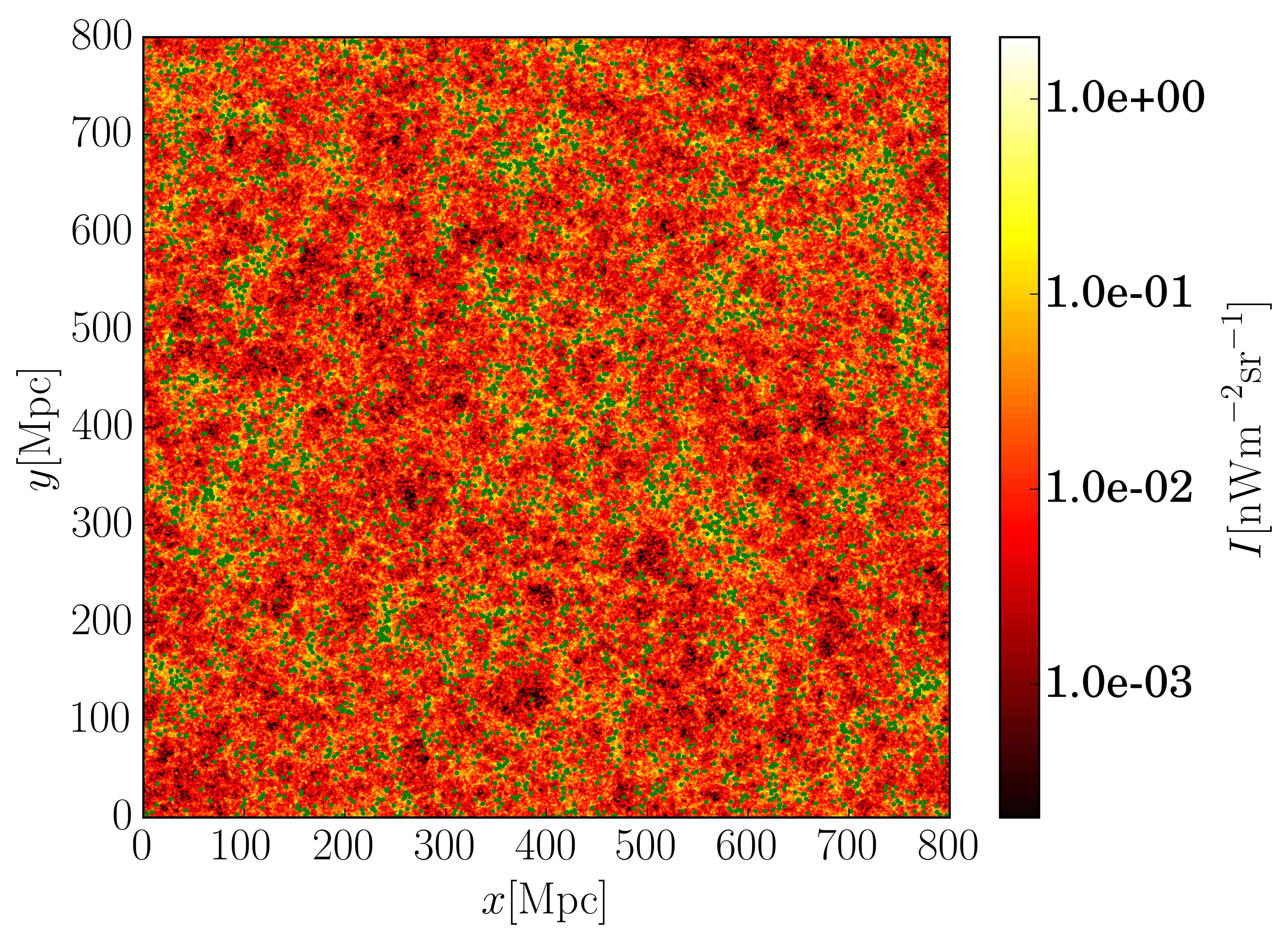}
}
\caption{{\bf Left}: diffuse \Lya emission in a slice at $z=6.6$. {\bf Right}: LAEs position in the same slice at $z = 6.6$ with $f_d = 0.1$. }
\label{fig:slices}
\end{figure*}
%
However, although promising, the feasibility of an IM experiment has not yet been fully demonstrated. The major challenge is that foregrounds dominate over line emission by several orders of magnitude. Although cleaning algorithms have been developed for, e.g. 21 cm radiation \citep{2006ApJ...650..529W, 2015arXiv150104429C, 2015arXiv150103823W}, it is not obvious that they are effective for other lines. Moreover some lines (such as \Lya and [CII]) present additional difficulties, such as line confusion (Comaschi et al. 2016, \cite{2014ApJ...785...72G}).

It is therefore quite possible that the first generation of intensity mappers will not be able to clean the signal up to a level sufficient to extract the targeted line power spectrum (PS). Thus it is crucial to study and develop alternative strategies, given that the first instruments are starting to be proposed \citep{2014arXiv1412.4872D, 2016arXiv160205178C, 2014SPIE.9153E..1WC}.
\cite{2015arXiv150404088C} recently showed that cross-correlating IM with point sources (specifically, QSOs in their case) can be very effective, even when only low quality data is available.  This approach has two important advantages: (i) the nature and redshift of the point sources are generally well known; (ii) one can observe both the cross-correlation and the auto-correlation of the point sources and use the latter to better constrain the diffuse line emission. Here we propose to combine \Lya IM with \Lya emitters (LAEs) data. This technique appears timely and it is motivated by the expectation that the next generation of LAEs surveys (such as the Subaru\footnote{\url{http://www.naoj.org/Projects/HSC/surveyplan.html}} ones) should find thousands of LAEs in the late EoR ($z = 5.7$ and $6.6$).
      
Beyond the low redshift work by \cite{2015arXiv150404088C}, this approach was previously investigated only for the 21 cm line at $z = 6.6$ and $7.3$. \cite{2016MNRAS.457..666V}  used N-body and radiative transfer simulations to compute the 21 cm-galaxy cross-power spectrum. They conclude that the Subaru Hyper Suprime Cam (HSC) and the LOw-Frequency Array for Radio astronomy (LOFAR) \citep{2013A&A...556A...2V} observations should show an anti-correlation on scales $k \approx 0.1 h {\rm Mpc}^{-1}$. \cite{2016arXiv160204837S} used seminumerical simulations to compute the 21cm-LAE cross-PS and cross-correlation function, studying different EoR scenarios and LAE duty cycles. 

In this work we study the properties of  the cross-correlation between diffuse \Lya emission and LAEs with seminumerical simulations \citep{2007ApJ...669..663M}. We use the model from \cite{2016MNRAS.455..725C} to populate with diffuse emission two $(800 {\rm Mpc})^3$ boxes at $z = 5.7$ and $6.6$. Next we stochastically generate the LAE population in several redshift slices. Each slice is comparable to a Subaru HSC observation both in terms of the number of LAEs it contains, and for its angular extension. In order to test the robustness of such observation to foreground residuals, we add a gaussian random noise with a variance up to 100 times the diffuse emission one and then try to recover the initial line intensity by comparing the measured cross-correlation function with the LAE auto-correlation one. We remark here that the present results depend only weakly on the specific line chosen, as the noise amplitude is scaled with the diffuse line variance.

The paper is organized as follows: in Sec. \ref{sec:method} we present our approach; Sec. \ref{sec:res} contains the results. Finally,  Sec. \ref{sec:conc} draws the conclusions. We assume a flat $\Lambda$CDM cosmology compatible with the latest Planck results ($h =0.67$, $\Omega_m = 0.32$, $\Omega_b = 0.049$, $\Omega_\Lambda = 1-\Omega_m$, $n = 0.97$, $\sigma_8 = 0.83$, \cite{2015arXiv150201589P}).

\section{Method}
\label{sec:method}

\begin{figure*}
\vspace{+0\baselineskip}
{
\includegraphics[width=0.45\textwidth]{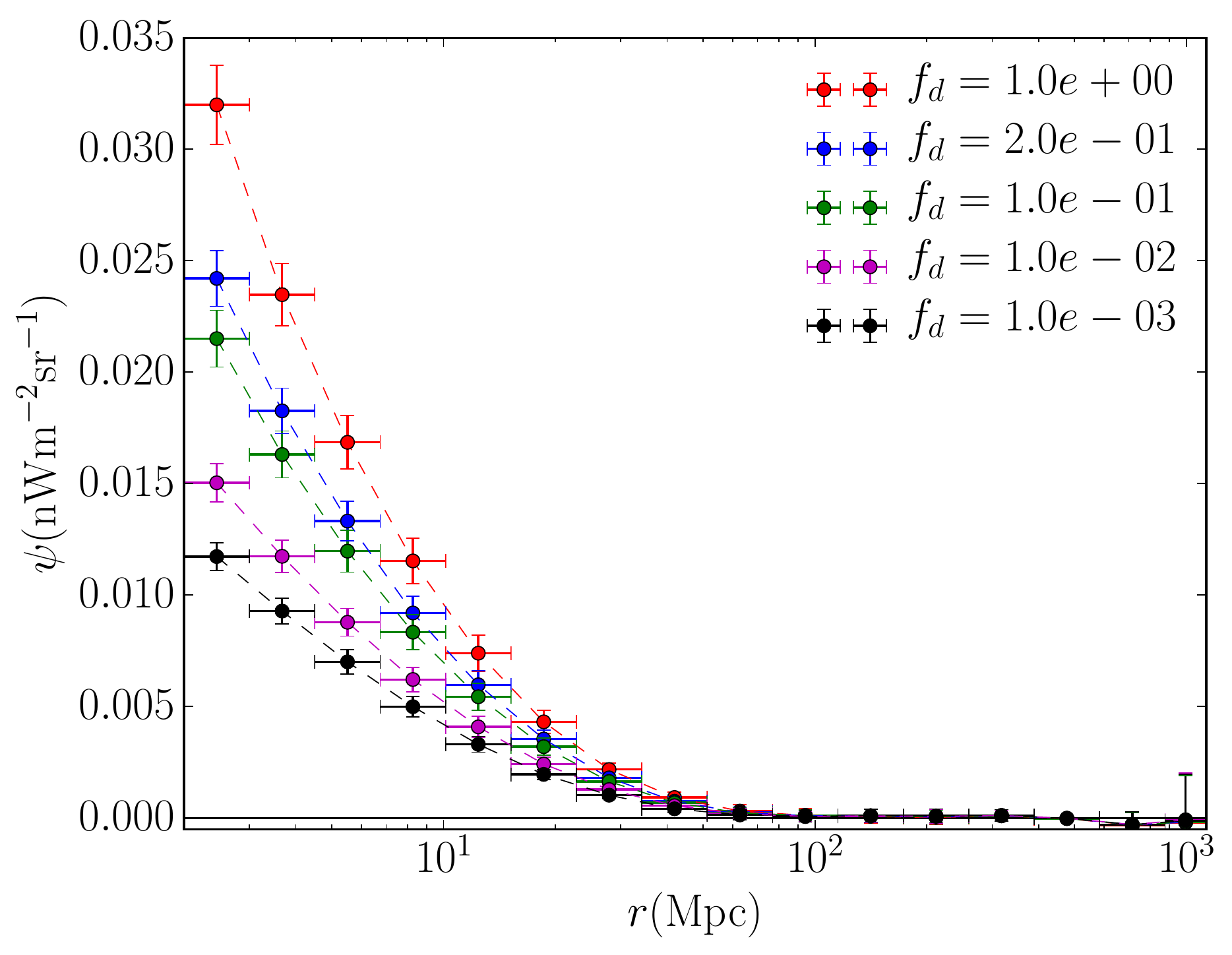}
\includegraphics[width=0.45\textwidth]{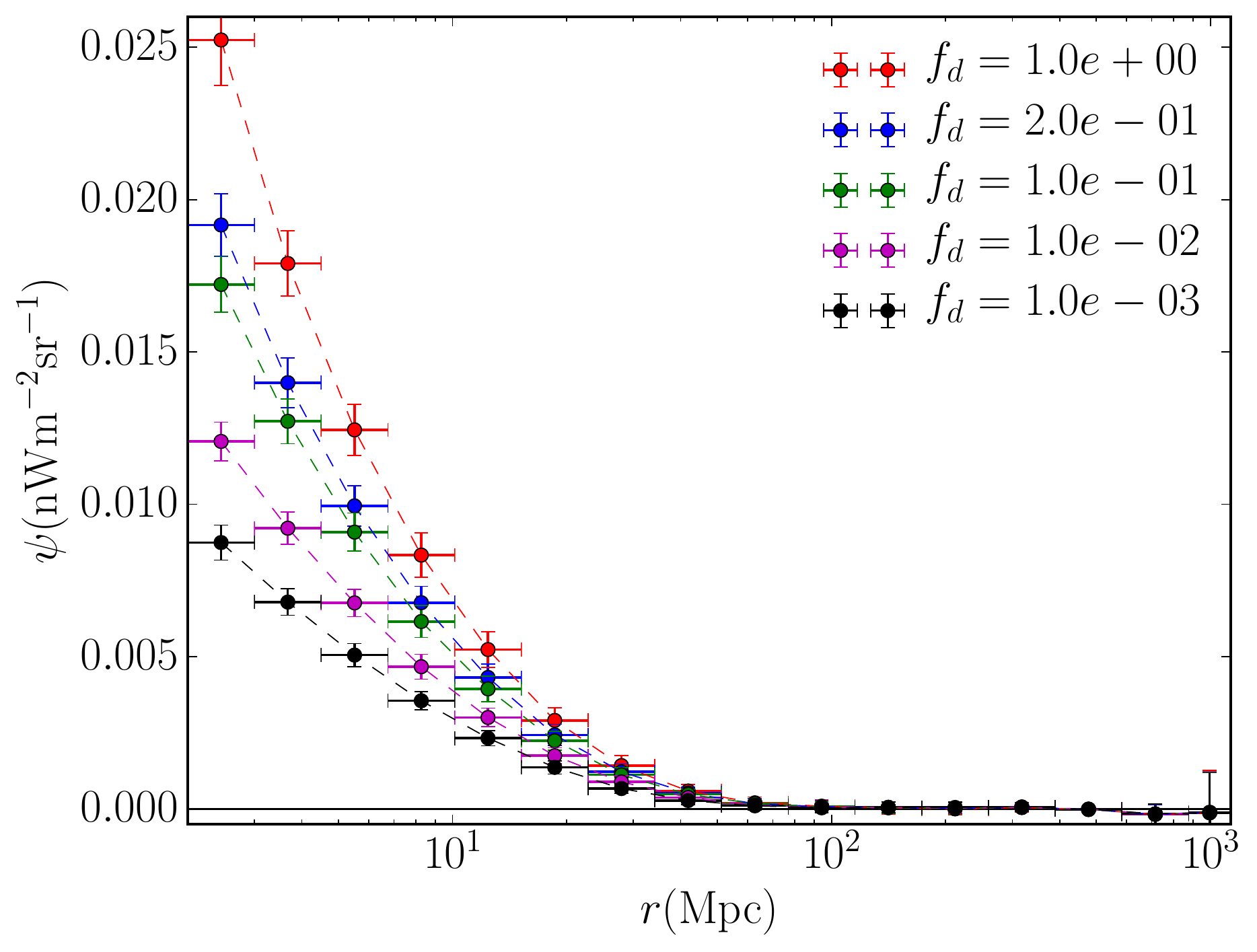}
}
\caption{Cross-correlation function between diffuse \Lya emission and LAE ($\cross$, eq. \eqref{psilyalae}) at $z = 5.7$ (left) and $z = 6.6$ (right) with $\Delta r = 1.5 r$. Different colors are for different duty cycles $f_d$.}
\label{fig:cflyaLAE}
\end{figure*}

We aim at deriving the cross-correlation between diffuse line emission and LAEs by producing mock observations with the code \texttt{DexM}\footnote{\url{http://homepage.sns.it/mesinger/Download.html}} \citep{2007ApJ...669..663M}. We generate two large ($800$ cMpc on a side) boxes at $z = 5.7$ and $z=6.6$ corresponding to $30.4(27.9){\rm deg}^2$ at $z=5.7(6.6)$, and resolve DM halos with minimum mass $M_{\rm min}(z = 5.7[6.6]) = 4.1[3.4]\times 10^{8}M_\odot$. The Subaru HSC survey has a redshift precision $\Delta z = 0.1$, corresponding to $\Delta l_z = 45.7(37.9)$ cMpc at $z = 5.7(6.6)$.  Therefore we divide each box in slices of thickness $\Delta l_z$ as above, obtaining $17$ (21) slices at $z = 5.7$ ($z = 6.6$).

In order to model diffuse line emission, we need to associate a line luminosity to each DM halo. We used the model by \cite{2016MNRAS.455..725C} (hereafter CF16). Although such method can be applied to any emission line, it is natural to consider IM in the \Lya line as LAE surveys are in the same band and share similar technical solutions (for example, a comparable angular resolution). Note, however, that the main conclusions of this work do not depend on the specific line used, for reasons discussed in Sec. \ref{sec:res}.

We compute the \Lya intensity map by dividing each slice in pixels of size $1$ Mpc (corresponding to $\Delta \theta = 24.8(23.8)$arcsec at $z = 5.7(6.6)$), considering only the \Lya emission from the ISM of galaxies. According to CF16 the IGM emission dominates the mean \Lya intensity; however, as we are here interested in fluctuations and IGM emission is very smooth on scales $k > 0.01 h{\rm Mpc}^{-1}$, we neglect such process. Fig. \ref{fig:slices} (left) shows the \Lya emission map of a representative slice at $z=6.6$. The mean \Lya intensity is $I_\lya = 3.51(1.99)\times 10^{-2} {\rm nW~}{\rm m}^{-2}{\rm sr}^{-1}$ at $z = 5.7(6.6)$, with a variance $\sigma_\alpha = 5.23(3.43)\times 10^{-2} {\rm nW~}{\rm m}^{-2}{\rm sr}^{-1}$. 

Next we populate the slices with LAEs. Since the LAE duty cycle ($f_d$, the fraction of galaxies that are classified as LAEs at a given time) is not well constrained, we considered five different values, $f_d =(1, 0.2, 0.1, 0.01, 0.001)$. We generate the LAE list stochastically among the galaxies in a given slice, fixing the total number of LAEs as $N_{\rm LAE} = 14000(5500)$ at $z=5.7(6.6)$. The value of $N_{\rm LAE}$ is consistent with the expected performance of the Subaru HSC survey (M. Ouchi, private communication).
We rank halos by mass and, starting from the most massive, we decide if a halo is a LAE randomly with probability $f_d$; we stop when we reach $N_{\rm LAE}$ extractions. 
Fig. \ref{fig:slices} (right) shows the LAE distribution in the same slice of Fig. \ref{fig:slices} (left), with $f_d = 0.1$. It is evident that both the diffuse \Lya emission and the LAE distribution trace the large scale DM distribution.

We do not include LAE emission in the diffuse \Lya emission. This is a good approximation because the bulk of diffuse \Lya radiation is dominated by LBG emission (e. g. \cite{2015arXiv150404088C}). CF16 is consistent with this hypothesis and it predicts the \Lya diffuse intensity independently of the observed LAE LF. Moreover, the LAE flux is likely to be will be lost anyway because pixels containing a LAE might be removed from the intensity map to better recover the signal from unresolved sources.

Finally, we compute the two-point correlation function between \Lya intensity and LAE with the same approach followed by \cite{2015arXiv150404088C}
\begin{equation}
\label{psilyalae}
  \cross(r, \Delta r) = \frac{1}{N_c} \sum_{i=1}^{N_c} \Delta I_\lya({\bf x}_i);
\end{equation} 
$N_c$ is the number of LAE-pixels pairs with distance between $r$ and $r+\Delta r$; $\Delta I_\lya({\bf x}_i)$ is the fluctuation of the \Lya intensity in the $i$-th pair pixel.

For comparison we computed also the two-point LAE auto-correlation function, i.e.
\begin{equation}
\label{psilaelae}
 \psi_{\rm LL}(r, \Delta r) = \frac{N_{c, l} - \langle N_{c, l} \rangle}{\langle N_{c, l} \rangle},
\end{equation}
where $N_{c, l}$ is the number of LAE pairs with distance between $r$ and $r + \Delta r$ and $\langle N_{c, l} \rangle$ in the expected value for a random LAE distribution.

\section{Results}
\label{sec:res}

Fig. \ref{fig:cflyaLAE} shows $\cross(r)$ for $z = 5.7$ and $6.6$ (computed considering bins with $\Delta r = 1.5 r$); the variance is the one among different slices. For all the five $f_d$ considered the amplitude of the correlation signal is sufficient to derive a model-dependent estimate of the diffuse line intensity. 

\begin{figure*}
\vspace{+0\baselineskip}
{
\includegraphics[width=0.45\textwidth]{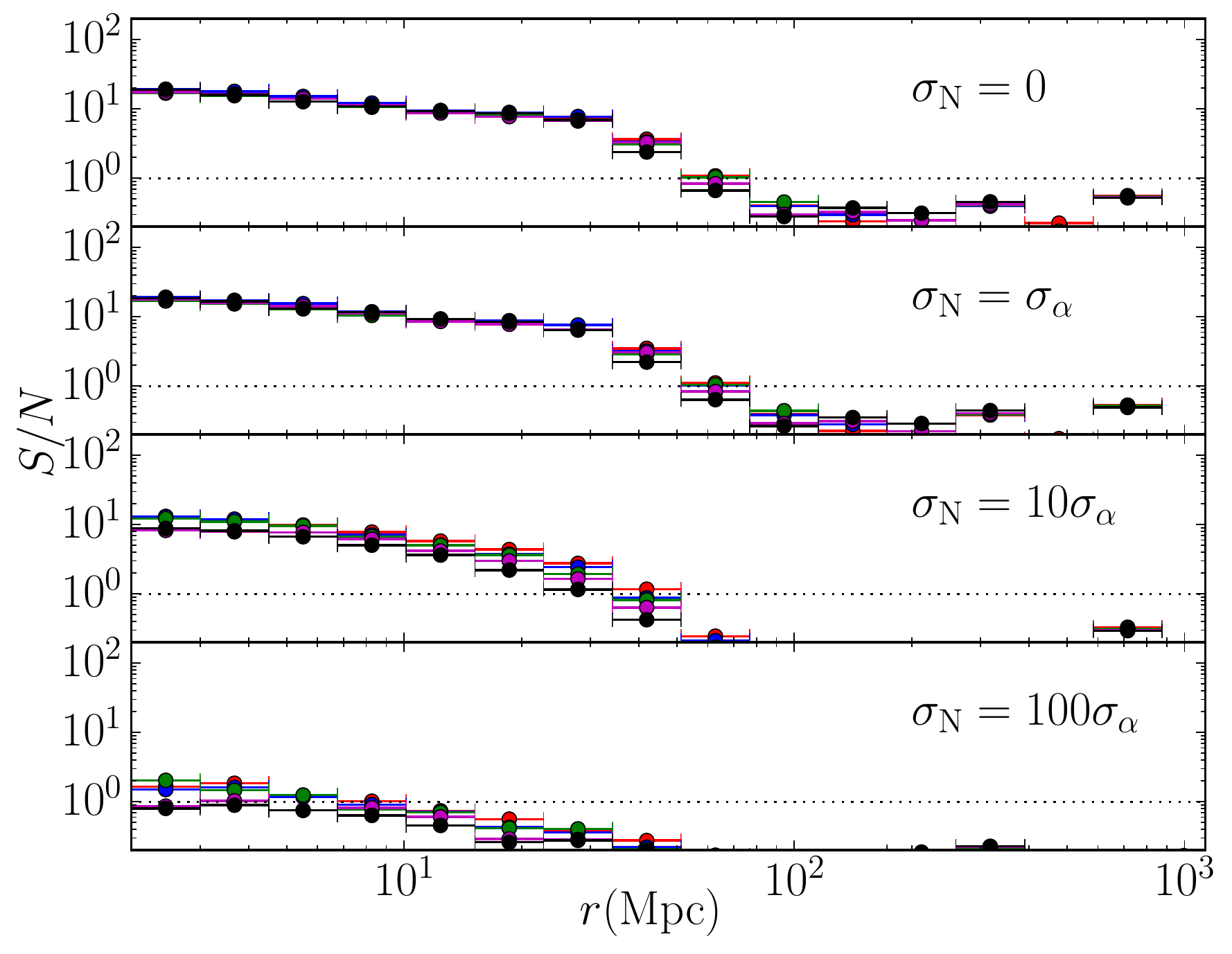}
\includegraphics[width=0.45\textwidth]{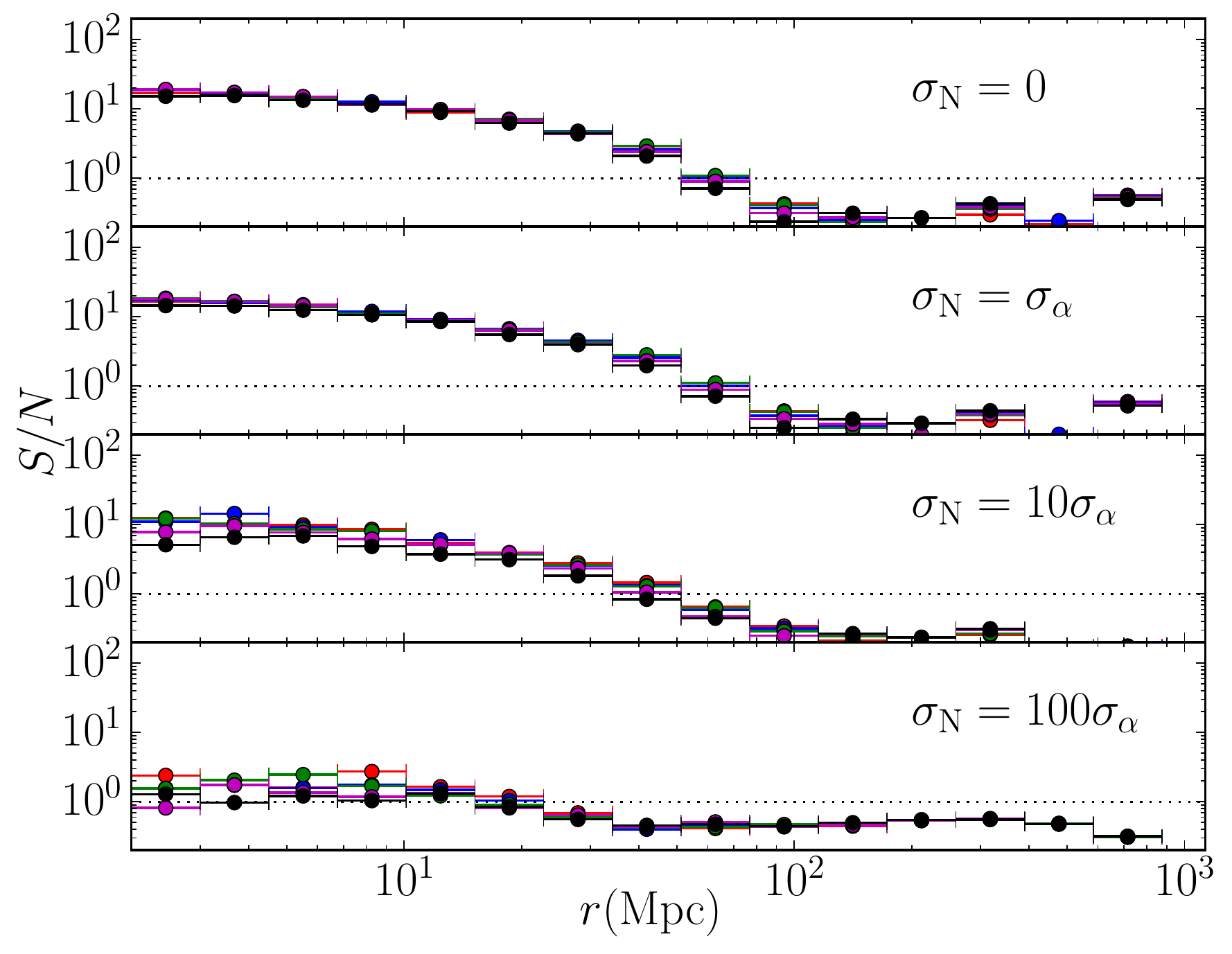}
}
\caption{S/N of the cross-correlation function in Fig. \ref{fig:cflyaLAE}. In the top panel no noise is included. In the lower panels a gaussian random noise is added to the \Lya fluctuations; the variance of the noise is $\sigma_{\rm N} = (0, 1, 10, 100)\sigma_\alpha$, from top to bottom. Different colors refer to different duty cycle values, $f_d$, as in Fig. \ref{fig:cflyaLAE}.}
\label{fig:sn}
\end{figure*}

A difficulty arises, though. A realistic intensity map will contain several sources of foregrounds that need to be removed before it is cross-correlated with LAEs. Because of this, a key step is to assess the robustness of the predicted cross-correlation signal to residual noise. To this aim  we superimpose a gaussian random noise with variance $\sigma_N = n \sigma_\alpha$, $n = (0, 1, 10, 100)$ to the \Lya intensity map, and compute again $\cross$. Fig. \ref{fig:sn} shows the dependence of the signal-to-noise (S/N) ratio on the noise. One can see that the Ly$\alpha$-LAE cross-correlation is very solid and can lead to model dependent estimates of the diffuse line intensity even with a residual noise with a variance $2$ orders of magnitude larger than the signal.

We point out that this result is almost independent of the particular line chosen for the diffuse emission. In fact, we have assumed a random noise variance purely proportional to the variance of the diffuse emission, without specifying at any level the actual nature of the noise. The only influence of the line choice on the results is the relative distribution of the emitted line luminosity on galaxy mass. This relation can influence both the bias and the shot-noise of line emission. However we do not expect that it could significantly affect the noise resilience and S/N of this cross-correlation.

To estimate the diffuse line intensity we proceed as follows. We assume that both $\cross(r)$ and $\psi_{LL}(r)$ are proportional to the dark matter two-point correlation function, $\psi_{\rm DM}(r)$:
\begin{gather}
  \cross(r) = \langle b \rangle_\lya I_\lya \langle b \rangle_{\rm LAE} \psi_{\rm DM}(r)\\
  \psi_{LL}(r) = \langle b \rangle_{\rm LAE}^2 \psi_{\rm DM}(r)
\end{gather}
where $\langle b \rangle_\lya$ is the \Lya luminosity weighted mean bias, and $\langle b \rangle_{\rm LAE}$ is the mean LAE bias. $\psi_{LL}(r)$ is computed using the estimator in eq. \eqref{psilaelae}.

Therefore
\begin{equation}
\label{irec}
 \frac{\cross(r)}{\psi_{LL}(r)} = \frac{\langle b \rangle_\lya}{\langle b \rangle_{\rm LAE}} I_\lya.
\end{equation}

We use eq. \eqref{irec} to estimate $I_\lya$ recovered from our mock observations, and compute $\langle b \rangle_\lya$ and $\langle b \rangle_{\rm LAE}$ from the CF16 model\footnote{The mass function in CF16 is slightly different from the one in the \texttt{DexM} code. We neglect this small inconsistency.}. 
\begin{figure}
\vspace{+0\baselineskip}
{
\includegraphics[width=0.45\textwidth]{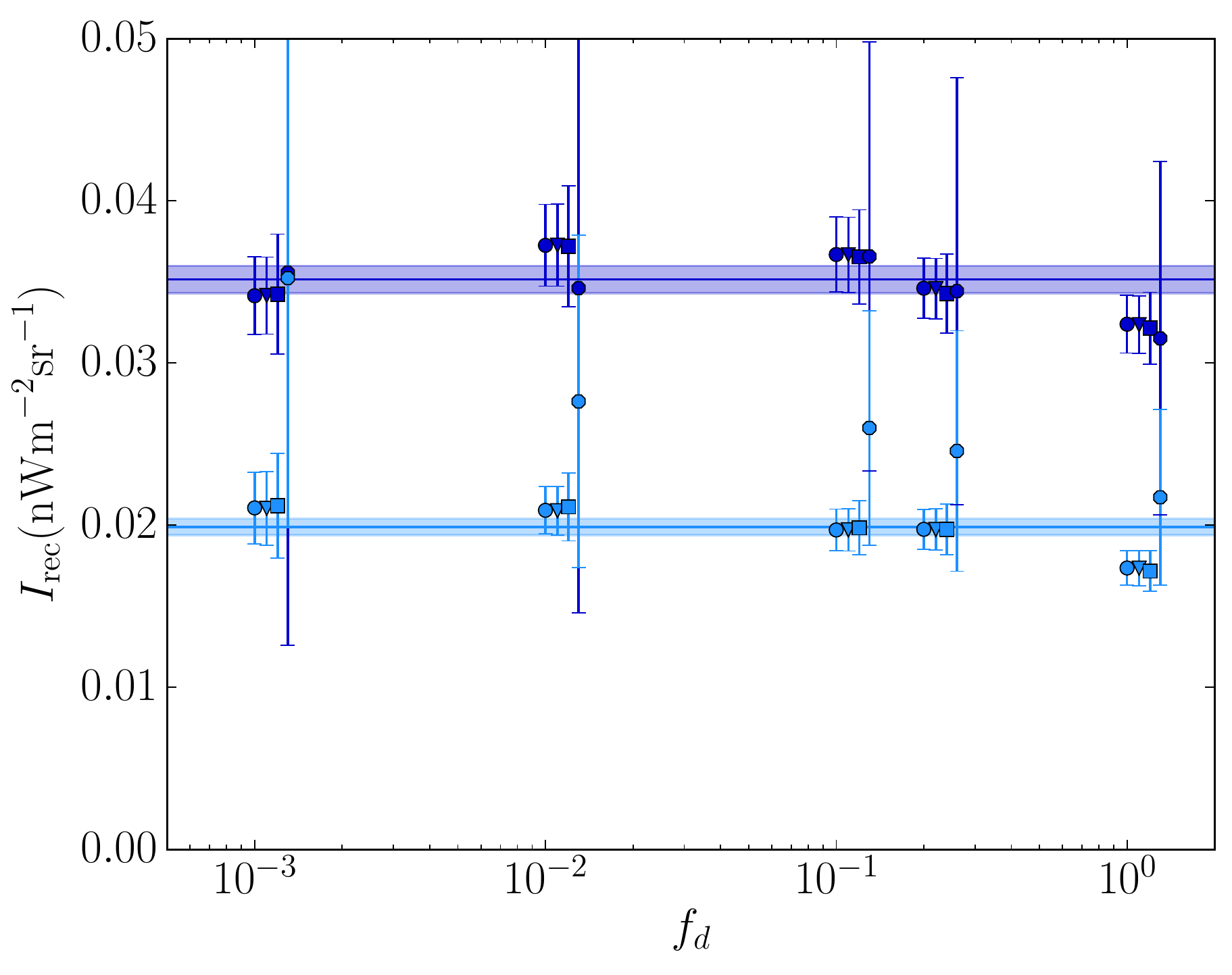}
}
\caption{Mean \Lya intensity recovered with eq. \eqref{irec}. Dark(Light) blue points are for $z=5.7(6.6)$, as a function of the LAE duty cycle $f_d$. At each $f_d$ four points show the results for different $\sigma_{\rm N}$: $\sigma_{\rm N} = (0, 1, 10, 100)\sigma_\alpha$ from left to right. The horizontal lines and the shaded region show the mean \Lya intensity in the mock slices and its 1-$\sigma$ variance.}
\label{fig:irec}
\end{figure}
The results (shown in Fig. \ref{fig:irec}) are in surprisingly good agreement with the actual $I_\lya$ values from the mock slices. Fig. \ref{fig:irec} shows that the Ly$\alpha$-LAE cross-correlation can be successfully used to determine the diffuse line intensity level and that this estimate is solid to foregrounds with a variance $\simlt 10$ times larger than the signal. Even if the variance is as large as $100 \sigma_\alpha$ our method can recover at least the order of magnitude of $I_\lya$.   
\section{Summary and Conclusions}
\label{sec:conc}
We have studied the cross-correlation $\cross(r)$ between diffuse \Lya line emission and the Subaru HSC \Lya emitters and its robustness to foregrounds. We computed the diffuse \Lya emission from dark matter halos in a $(800{\rm Mpc})^3$ box at $z = 5.7$ and $6.6$. We divided the box in slices and populate them stochastically with $14000(5500)$ LAEs at $z = 5.7(6.6)$, considering duty cycles from $10^{-3}$ to $1$; both the LAE number density and the size of the slices are consistent with the expected outcome of the Subaru HSC survey. We added gaussian random noise with variance $\sigma_{\rm N}$ up to 100 times the variance of the \Lya emission $\sigma_\alpha$ to simulate foregrounds and compute $\cross(r)$.

We found that the signal-to-noise of the observed $\cross(r)$ did not change significantly if $\sigma_{\rm N} \le 10 \sigma_\alpha$ and we have showed that in these conditions the mean line intensity $I_\lya$ can be recovered independently of the LAE duty cycle. Even if $\sigma_{\rm N} = 100 \sigma_\alpha$, $I_\alpha$ could be constrained within a factor $2$. We point out that these results depend only very weakly on the line chosen to perform the intensity mapping experiment.

These results are very promising because they show that even with a dirty IM observation it is possible to recover the diffuse line intensity by cross-correlating with point sources, such as LAE. Since removing the residual foregrounds (mostly coming from interloping lines)  in future IM surveys will not be an easy task, relying on a solid cross-correlation is crucial to extract information from IM experiments. In addition, the cross-correlation with point sources can provide an independent test of the quality of the recovered line signal after foreground removal. 

These results represent only a first step towards a more comprehensive feasibility study of the proposed technique. Nevertheless they clearly demonstrate the potentiality of the approach. Future work should be devoted to simulate mock observations with realistic foreground residuals including interlopers, and more sophisticated LAE models. The estimate of the diffuse line intensity might also benefit by a more advanced treatment of all the contributing processes. These improvements are certainly encouraged by the success of the simple model presented here. 
\bibliography{paper_LAE}

\end{document}